\pdfoutput=1
\documentclass[12pt,titlepage]{article}

\font\grande=cmr9.5 scaled \magstep4
\font\medio=cmr9.5 scaled \magstep2
\outer\def\beginsection#1\par{\medbreak\bigskip
      \message{#1}\leftline{\bf#1}\nobreak\medskip
\vskip-\parskip
      \noindent}
\usepackage{graphicx} 
\begin{document}
\bibliographystyle {unsrt}

\titlepage

\begin{flushright}
\end{flushright}

\vspace{1cm}
\begin{center}
{\grande Stimulated emission of relic gravitons}\\
\vspace{0.5cm}
{\grande and their super-Poissonian statistics}\\
\vspace{1cm}
 Massimo Giovannini 
 \footnote{Electronic address: massimo.giovannini@cern.ch} \\
\vspace{1cm}
{{\sl Department of Physics, CERN, 1211 Geneva 23, Switzerland }}\\
\vspace{0.5cm}
{{\sl INFN, Section of Milan-Bicocca, 20126 Milan, Italy}}
\vspace*{1cm}
\end{center}

\vskip 0.3cm
\centerline{\medio  Abstract}
\vskip 0.1cm
The degree of second-order coherence of the relic gravitons produced  from the vacuum 
is super-Poissonian and larger than in the case of a chaotic source characterized by 
a Bose-Einstein distribution. If the initial state does not minimize the tensor Hamiltonian
and has a dispersion smaller than its averaged multiplicity, the overall statistics is  by definition
sub-Poissonian. Depending on the nature of the sub-Poissonian initial state, 
the final degree of second-order coherence of the quanta produced by stimulated emission may diminish 
(possibly even below the characteristic value of a chaotic source)  but it always remains larger than one 
(i.e. super-Poissonian). When the initial statistics is Poissonian (like in the case of a coherent state or for 
a mixed state weighted by a Poisson distribution) the degree of second-order coherence of the produced 
gravitons is still super-Poissonian. Even though the quantum origin of the relic gravitons inside the Hubble radius 
can be effectively disambiguated by looking at the corresponding Hanbury Brown-Twiss correlations, 
the final distributions caused by different initial states maintain their super-Poissonian character 
which cannot be altered.
\noindent

\vspace{5mm}
\vfill
\newpage
The current lore suggests that the seeds of large-scale inhomogeneities have a quantum mechanical origin as speculated by Sakharov \cite{sakharov} more than fifty years ago. Assuming this exciting possibility is indeed realized in nature, so far only the effects of curvature phonons have been indirectly observed through their imprint on the temperature and polarization anisotropies  of the Cosmic Microwave Background (CMB). The curvature phonons associated with the scalar fluctuations of the geometry should be complemented by the relic gravitons whose production has been suggested on a general ground \cite{GR} even prior to the formulation of conventional inflationary scenarios \cite{INF}.  The combined analyses of the large-scale data sets suggest that at a typical pivot scale ${\mathcal O}(0.002)\, \mathrm{Mpc}^{-1}$ the tensor power spectrum must be conservatively smaller than one hundredth of the scalar power spectrum (see e.g. \cite{RT}). While current observations  are hoping for potential signals of the large-scale tensor fluctuations in the polarization anisotropies of the CMB, the wide-band interferometers are now exploring much shorter length-scales where, according to the common hopes, the primeval gravitational waves should be hopefully detected during the forthcoming decades \cite{INTER}. While the typical frequency of current CMB experiments corresponds to few aHz ($1\, \mathrm{aHz} = 10^{-18}\, \mathrm{Hz}$) the wide-band interferometers are today operating between few Hz and $10$ kHz.

The transverse and traceless fluctuations of the metric are conventionally denoted\footnote{We shall consider hereunder a conformally flat background geometries whose metric can be expressed as $\overline{g}_{\mu\nu} = a^2(\tau) \eta_{\mu\nu}$ where $\eta_{\mu\nu}$ is the Minkowski metric with signature $(+,\, -,\,-,\,-)$ and $a(\tau)$ is the scale factor in the conformal time parametrization. Standard notations will be used throughout and the Hubble rate will be given by $H= {\mathcal H}/a$ where ${\mathcal H} = a^{\prime}/a$ and the prime denotes a derivation with respect to the conformal time coordinate $\tau$.}
by $\delta_{t} g_{ij} = - a^2 h_{ij}(x)$ where $x_{i} \equiv (\vec{x}_{i}, \, \tau_{i})$; for convenience 
we shall be working in terms of the rescaled tensor amplitude $\hat{\mu}_{ij}(x) = \hat{h}_{ij}(x) \,a(\tau)$ where the hats will denote throughout the quantum field operators whose explicit expressions are: 
\begin{equation}
\hat{\mu}_{ij}(x) = \frac{\sqrt{2} \ell_{P}}{( 2\pi)^{3/2}} \sum_{\lambda} \int d^{3} k \,e^{(\lambda)}_{ij} \, \hat{\mu}_{\vec{k}\, \lambda}(\tau) e^{- i \vec{k} \cdot\vec{x}}, \qquad \hat{\mu}_{\vec{k}\, \lambda}(\tau)= \frac{\hat{a}_{\vec{k}\,\lambda} + \hat{a}^{\dagger}_{\vec{-k}\,\lambda}}{\sqrt{2 k}},
\label{exp1}
\end{equation}
where $\ell_{P} = \sqrt{8 \pi G}$ and $\lambda= (\oplus, \otimes)$ runs over the two polarizations of the graviton. The evolution of the background geometry acts in fact as a parametric amplifier ultimately producing pairs of gravitons with opposite momenta according to the Hamiltonian:
\begin{equation}
\hat{H} = \frac{1}{2} \int d^{3} p \sum_{\lambda} \biggl\{ p \biggl[ \hat{a}^{\dagger}_{\vec{p}\,\lambda} \hat{a}_{\vec{p}\,\lambda} + \hat{a}_{- \vec{p}\,\,\lambda} \hat{a}^{\dagger}_{-\vec{p}\,\,\lambda} \biggr] + i {\mathcal H} \biggl[ \hat{a}^{\dagger}_{-\vec{p}\,\,\lambda} \hat{a}_{\vec{p}\,\,\lambda}^{\dagger} 
-  \hat{a}_{\vec{p}\,\lambda} \hat{a}_{-\vec{p}\,\,\lambda}\biggr] \biggr\},
\label{H1}
\end{equation}
where $\hat{a}_{\vec{k}\,\lambda}$ and 
$\hat{a}^{\dagger}_{\vec{k}\,\lambda}$ are the annihilation and creation operators of a graviton with comoving three-momentum 
$\vec{k}$ and polarization $\lambda$; they obey the standard commutation relations $[\hat{a}_{\vec{k}\,\alpha},\, \hat{a}^{\dagger}_{\vec{p}\,\beta}] = \delta^{(3)}(\vec{k} - \vec{p}) \delta_{\alpha\beta}$. Equation (\ref{H1}) can be diagonalized via a canonical transformation of the following type 
\begin{eqnarray}
\hat{a}_{\vec{p}\,\,\alpha}(\tau) &=& u_{p\,\,\alpha}(\tau,\tau_{i}) \,\hat{b}_{\vec{p}\,\,\alpha}(\tau_{i}) -  
v_{p\,\,\alpha}(\tau,\tau_{i}) \,\hat{b}_{-\vec{p}\,\,\alpha}^{\dagger}(\tau_{i}),
\label{AA10}\\
\hat{a}_{-\vec{p}\,\,\alpha}^{\dagger}(\tau) &=& u_{p\,\,\alpha}^{*}(\tau,\tau_{i})\, \hat{b}_{-\vec{p}\,\,\alpha}^{\dagger}(\tau_{i})  -  v_{p\,\,\alpha}^{*}(\tau,\tau_{i})\, \hat{b}_{\vec{p}\,\,\alpha}(\tau_{i}).
\label{AA11}
\end{eqnarray}
Thanks to Eqs. (\ref{AA10}) and (\ref{AA11})  the evolution of the creation and annihilation operators can be rephrased in terms of the two complex functions $u_{p\,\,\alpha}(\tau,\tau_{i})$ and $v_{p\,\,\alpha}(\tau,\tau_{i})$ that must satisfy the condition $|u_{p\,\,\alpha}(\tau,\tau_{i})|^2 - |v_{p\,\,\alpha}(\tau,\tau_{i})|^2 =1$ (following from the unitary evolution of the system). The complex functions of Eqs. (\ref{AA10}) and (\ref{AA11}) must therefore depend, 
for each polarization and for a given three-momentum, upon three real functions, i.e. two phases and one amplitude. 
Consequently, without loss of generality, we can posit $u_{k,\,\alpha}(\tau,\tau_{i}) = e^{- i\, \delta_{k,\,\alpha}} \cosh{r_{k,\,\alpha}}$ and $v_{k,\,\alpha}(\tau,\tau_{i}) =  e^{i (\theta_{k,\,\alpha} + \delta_{k,\,\alpha})} \sinh{r_{k,\,\alpha}}$.

The adiabatic and Gaussian nature of the initial conditions of the Einstein-Boltzmann hierarchy
suggested by the current large-scale data \cite{RT}  is compatible with a quantum mechanical 
origin of the curvature perturbations but the observed  patterns of the temperature and polarization 
anisotropies are, per se, not sufficient to establish their quantum origin beyond any reasonable doubt. 
A well defined physical question concerns therefore the origin of the observed fluctuations and it
 would be relevant to develop a set of sufficient criteria reducing the vague problem of the quantumness 
 of the fluctuations to a more empirical strategy aimed at a specific analysis of their correlations. 
One of the few possibilities along this direction relies on the application of 
Hanbury Brown-Twiss interferometry \cite{HBT0} to the large-scale curvature inhomogeneities and 
to relic gravitons \cite{HBT1}. In a nutshell the idea of Hanbury Brown-Twiss (HBT) interferometry
is that not only amplitudes can interfere (as in the case of conventional Young-type experiments)
but also intensities. Consequently the logic of the HBT interferometry is deeply rooted in the analysis 
of intensity correlations as stressed for the first time by Glauber and Sudarshan \cite{glauber}.
The current observations of gravitational waves by wide-band interferometers  between few Hz and $10$ kHz 
are only sensitive to the average multiplicity of the gravitons and hence to their degree of first-order coherence. 
This is true even assuming, rather optimistically, that the current interferometers will be one day sufficiently 
sensitive to detect a background of relic gravitons. If the initial state of the relic gravitons is the vacuum, the degree of second-order coherence can be computed within a consistent field-theoretical description and the results are \cite{HBT1}
\begin{equation}
g^{(2)}(x_{1}, x_{2}) \to \frac{41}{30}, \qquad \overline{g}^{(2)}(x_{1}, x_{2}) \to 3,
\label{three}
\end{equation}
where $g^{(2)}$ refers to the case where the sum over the polarizations is 
consistently implemented\footnote{Note that in a previous analysis (see \cite{HBT1}, third paper) $g^{(2)}(x_1, x_{2})$ has been estimated as $71/60\simeq 1.18$. A more accurate analysis (see last paper of Ref. \cite{HBT1})  shows that this result must be corrected as $41/30 \simeq 1.36$.} in the definition of the intensity of the relic gravitons; $\overline{g}^{(2)}$ denotes the  degree of second-order coherence computed from a single polarization of the gravitons. The results of Eq. (\ref{three}) show that the relic gravitons are always super-Possonian but the effect of the polarizations is a progressive reduction of the degree of second-order coherence. This reduction preserves the super-Poissonian character of the quantum state so that the Poissonian limit is never reached \cite{HBT1}.
The results of Eq. (\ref{three}) assumed an initial state minimizing the tensor Hamiltonian. In this paper 
we shall instead be concerned with the complementary situation where the initial state is not the vacuum.

The degree of second-order coherence $\overline{g}^{(2)}$ of Eq. (\ref{three}) coincides with the results obtained by considering a single mode of the field (i.e. by neglecting the dependence of the creation and annihilation operators upon the comoving three-momentum and upon the polarization) as it happens, in some cases, for the optical photons in a cavity \cite{QO}.  In the present context the single-mode limit actually accounts for the statistics of the graviton pairs:  if the operator $\hat{c}$ annihilates a graviton of comoving three-momentum $\vec{k}$ while 
$\hat{d}$ describes the annihilation of a graviton of opposite comoving three-momentum (i.e. $- \vec{k}$)
we have that, according to Eq. (\ref{AA10}):
\begin{equation}
\hat{a}_{1} = e^{- i \delta} \cosh{r} \,\hat{c} - e^{i (\theta+ \delta)} \sinh{r}\, \hat{d}^{\dagger},\qquad 
\hat{a}_{2} = e^{- i \delta} \cosh{r} \,\hat{d} - e^{i (\theta+ \delta)} \sinh{r} \, \hat{c}^{\dagger}.
\label{degA33b}
\end{equation}
But since the two oscillators obey the standard commutation relations (i.e. $[\hat{c}, \hat{c}^{\dagger}] =1$ and $[\hat{d}, \hat{d}^{\dagger}] = 1$) and  are also  mutually commuting (i.e. $[\hat{c}, \hat{d}]=0$) the sum $\hat{a} = (\hat{a}_{1} + \hat{a}_{2})/\sqrt{2}$ obeys the standard commutation relation of a single mode (i.e. $[\hat{a}, \hat{a}^{\dagger}] =1$). Consequently, thanks to Eq.  (\ref{degA33b}), the averaged multiplicity is the sum of the individual multiplicities of the gravitons with opposite momenta, i.e. $\langle z| \hat{a}^{\dagger} \hat{a} | z\rangle= (\langle z| \hat{c}^{\dagger} \hat{c} | z\rangle + \langle z| \hat{d}^{\dagger} \hat{d} | z\rangle)/2 = \sinh^2{r}$.  In the single-mode limit the degree of second-order coherence of the relic gravitons can be expressed as\footnote{The average multiplicity of the gravitons shall be denoted by 
$\langle \hat{n} \rangle = {\mathrm Tr}[ \hat{\rho} \, \hat{a}^{\dagger}\,\hat{a}]$ while  
$\sigma^2 = \langle \hat{n}^2 \rangle - \langle \hat{n} \rangle^2$ is the dispersion. With these 
notations the second equality in Eq. (\ref{one}) immediately follows.}:
\begin{equation}
g^{(2)}_{s} = \frac{{\mathrm Tr}[ \hat{\rho} \, \hat{a}^{\dagger} \, \hat{a}^{\dagger} \, \hat{a} \,\hat{a} ]}{ \{{\mathrm Tr}[ \hat{\rho} \, \hat{a}^{\dagger}\,\hat{a}] \}^2} = 1 + \frac{\sigma^2 - \langle \hat{n}\rangle }{ \langle \hat{n}\rangle^2},
\label{one}
\end{equation}
where the subscript {\em s}  refers to the single-mode approximation while $\hat{\rho}$ 
denotes the density operator: indeed the states involved in the present discussion can either be pure or mixed.  
When the dispersion coincides with the mean value (i.e. $\sigma^2 = \langle \hat{n} \rangle$) as in the case of a Poisson distribution, then $g_{s}^{(2)} \to 1$ and this limit is reached by a single-mode coherent 
 state (and it is commonly referred to as the Poisson limit \cite{QO}). The state of the relic gravitons in the single-mode approximation can be schematically written as\footnote{The operator $S(z)$ is the so-called squeezing operator and $R(\delta)$ is the rotation operator \cite{QO}.}
$| z\, \delta \rangle = S(z) \, R(\delta) \, |\mathrm{vac} \rangle$ where $S(z)= \exp{[(z^{*}/2) \hat{a}^2 - (z/2) \hat{a}^{\dagger\, 2}]}$ and 
$R(\delta) = \exp{[ - i (\delta/2) \hat{a}^{\dagger} \hat{a}]}$. If the relic gravitons are parametrically amplified from the vacuum the degree of second-order coherence computed from Eq. (\ref{one}) for the state $| z\, \delta \rangle$ does not depend on $\delta$ and $\theta$ but only on the average multiplicity: 
\begin{equation}
g_{s}^{(2)} = 3 + \frac{1}{\overline{n}_{sq}}, \qquad \lim_{\overline{n_{sq}} \gg 1} g_{s}^{(2)} \to  3,
\label{twoa}
\end{equation}
where $\overline{n}_{sq} = \sinh^2{r}$ is the average multiplicity of the squeezed vacuum state. As anticipated the result (\ref{twoa}) agrees with Eq. (\ref{three}) since $\overline{g}^{(2)} = g_{s}^{(2)}$.

In the context of conventional inflationary models the initial state of the fluctuations 
is immaterial unless the total number of efolds is close to a certain 
critical number  $N_{crit} = {\mathcal O}(66)$; this figure can be derived by demanding that 
the inflationary event horizon redshifted at the present epoch coincides with the the Hubble radius today 
and, for this 
reason, the critical number of efolds\footnote{The value of $N_{crit}$ also depends on the post-inflationary history and onservative estimates suggest $N_{crit} = 63 \pm15$ \cite{NN}. In the case of a standard  post-inflationary history $N_{crit} = 63.6 + (1/4) \ln{\epsilon}$, where $\epsilon$ denotes the standard slow-roll parameter. According to some, for the consistency of the inflationary scenarios we must anyway demand that the total number of efolds greatly exceeds $N_{crit}$.} depends on the post-inflationary thermal history \cite{NN}.  With this specification the simplest non-trivial possibility beyond the vacuum is to consider a Fock state that is subsequently rotated [through $R(\delta)$] and then squeezed [through $S(z)$]. In the single-mode approximation the final state can then be schematically denoted by
$| z\, \delta\, n \rangle = S(z) R(\delta) | n \rangle$ and its associated degree of second-order coherence follows 
from Eq. (\ref{one}):
\begin{equation}
g_{s}^{(2)} = 1 - \frac{n ( 2 \overline{n}_{sq} + 1)}{[ (n+1) \overline{n}_{sq} + n (\overline{n}_{sq} +1)]^2}
+ \frac{\overline{n}_{sq} [ 2 n^2 (\overline{n}_{sq} + 1) + 2 n (\overline{n}_{sq} +1 ) + 2 \overline{n}_{sq} +1]}{[ (n+1) \overline{n}_{sq} + n (\overline{n}_{sq} +1)]^2},
\label{IN1}
\end{equation}
where $n$ denotes the  average multiplicity of the initial Fock state
while $\overline{n}_{sq} = \sinh^2{r}$ is the average multiplicity of the produced gravitons. When $\overline{n}_{sq} \ll n $ the degree of second-order coherence of the Fock state is recovered and it can only be smaller than $1$ (i.e. $g_{s}^{(2)} < 1$) \cite{QO}. From the viewpoint of the relic gravitons the limit $\overline{n}_{sq} \ll n $ is actually unphysical  since the average multiplicity of the created quanta always exceeds the multiplicity of the initial state (i.e. $\overline{n}_{sq} \gg n$). In the physical limit the initial Fock state can however reduce the final degree of second-order coherence. Indeed, when $\overline{n}_{sq}\gg 1 $ (for a fixed but otherwise arbitrary value of 
$n$), Eq. (\ref{IN1}) implies that\footnote{Note that the second limit in Eq. (\ref{IN2}) holds when $\overline{n}_{sq} > n$ and, at the same time, $n \gg 1$.}:
\begin{equation}
\lim_{\overline{n}_{sq} \gg 1} g_{s}^{(2)} = 1 + \frac{ 2 (n^2 + n + 1)}{( 2 n + 1)^2 } + {\mathcal O}\biggl(\frac{1}{n\,\overline{n}_{sq}}\biggr) \to \frac{3}{2}.
\label{IN2}
\end{equation}
 Equations (\ref{IN1}) and (\ref{IN2}) show that the degree of second-order coherence can
be reduced from $3$ (as in the case of a squeezed vacuum state of Eqs. (\ref{three}) and (\ref{twoa})) to $1.5$ 
provided the initial state is appropriately selected and assuming that the total number of inflationary efolds is sufficiently close to its critical value. By setting $\overline{n}_{sq}\to 0$ in Eq. (\ref{IN1}) we can verify that the degree of second-order coherence is sub-Poissonian \cite{QO} as implied by the Fock state in the absence of squeezed contribution.

A reduction of the degree of second-order coherence, analog to the one previously discussed, may also arise when the initial state of the quantum mechanical evolution is not pure (as in the case of Eqs. (\ref{IN1})--(\ref{IN2})) but mixed. Mixed states 
do not generally minimize the tensor Hamiltonian of Eq. (\ref{H1}) and are specified by assigning a density matrix $\hat{\rho} = \sum_{n} p_{n} |n \rangle \langle n|$ whose explicit form (in the Fock basis) depends on certain statistical weights $p_{n}$ with $\sum_{n} p_{n} = 1$ (in what follows all the sums over $n$ range between $n =0$ and $n \to \infty$). Instead of working with the probabilities $p_{n}$ it is more practical to deal with the corresponding characteristic functions $P(s) = \sum_{n} s^{n} p_{n}$ (obviously implying $P(1)=1$). The degree of second-order coherence of Eq. (\ref{one}) can then be computed in general terms and it depends on $\overline{n}_{sq}$ and on two supplementary parameters, namely $\overline{n}_{in} = \sum_{n} n\, p_{n} $ (i.e. the average multiplicity of the initial state) and $\overline{n^2_{in}} = \sum_{n} n^2 \, p_{n}$ (i.e. the second moment of the initial distribution): 
\begin{equation}
g_{s}^{(2)} = \frac{\overline{n^{2}_{in}} \,[ 6 \overline{n}_{sq}^2 + 6 \overline{n}_{sq} +1] }{[\overline{n}_{sq} (\overline{n}_{in} +1) + 
\overline{n}_{in}(\overline{n}_{sq} +1)]^2} 
+ \frac{\overline{n}_{in} [ 6 \overline{n}_{sq}^2 + 2 \overline{n}_{sq} -1] + \overline{n}_{sq} ( 3 \overline{n}_{sq} +1) }{[\overline{n}_{sq} (\overline{n}_{in} +1) + \overline{n}_{in}(\overline{n}_{sq} +1)]^2},
\label{IN3}
\end{equation}
A pretty useful parametrization of the characteristic function $P(s)$ is given by: 
\begin{equation}
P(s) = \frac{k^{k}}{[ k + \overline{n}_{in}(1 -s)]^{k}}, \qquad P(1) = 1, \qquad k> 0.
\label{IN3a}
\end{equation}
When $k =1$, $P(s)$ coincides with the characteristic function of the Bose-Einstein distribution 
[i.e. $p_{n} = \overline{n}_{in}^{n}/(\overline{n}_{in} +1)^{n +1}$]. When $k$ is integer Eq. (\ref{IN3a}) 
is related to a collection of Bose-Einstein distributions \cite{QO}. If $k \to \infty$ then $P(s)\to e^{\overline{n}_{in}(s-1)}$ 
 by so coinciding with the characteristic function of a Poisson 
distribution\footnote{From Eq. (\ref{IN3}) when $\overline{n}_{sq} \to 0$ the degree of second-order coherence goes as 
$g^{(2)}_{s} \to (1 + 1/k)$, i.e. $g_{s}^{(2)} \to 2$ for $k \to 1$, as expected in the case of a thermal mixture. In fact, that according to Eq. (\ref{IN3a}) $\overline{n^{2}_{in}} = [ \overline{n}_{in} + \overline{n}_{in}^2 (1 + 1/k)]$; to avoid potential ambiguities at the notational level it should be remarked that 
$\overline{n^{2}_{in}} \neq \overline{n}_{in}^2$ since $\overline{n^{2}_{in}}$ is the second-order momentum 
of $p_{n}$ while  $\overline{n}_{in}^2$ is the square of the average.  }. For $\overline{n}_{sq} \gg \overline{n}_{in} \gg 1$, Eq. (\ref{IN3}) 
implies that 
\begin{equation}
\lim_{\overline{n}_{sq} \gg \overline{n}_{in} \gg 1} \overline{g}_{s}^{(2)} = \frac{3}{2} \biggl(1 + \frac{1}{k}\biggr) + {\mathcal O}\biggl(\frac{1}{\overline{n}}_{in}\biggr) + {\mathcal O}\bigg( \frac{1}{ \overline{n}_{in} \overline{n}_{sq}^2}\biggr)+ 
 {\mathcal O}\bigg( \frac{1}{ \overline{n}_{sq} \overline{n}_{in}^2}\biggr).
\label{IN4}
\end{equation}
According to Eq. (\ref{IN4}) $g_{s}^{(2)}\to 3$ in the Bose-Einstein case (i.e. $k \to 1$); this result basically coincides with Eqs. (\ref{three}) and (\ref{twoa}). We conclude that an initial thermal mixture is unable to decrease the degree of second-order coherence of the squeezed vacuum. The opposite is true if $k > 1$ since, in this case, $g_{s}^{(2)} < 3$. The minimal degree of second-order coherence in  Eq. (\ref{IN4}) (i.e. $g^{(2)}_{s} \to 3/2$)
follows in the limit $k \to \infty$ where the initial (mixed) state is characterized exactly by a  Poisson distribution. The coincidence between this result and the numerical value obtained for a Fock state (see Eqs. (\ref{IN1})--(\ref{IN2})) does not seem accidental since, for an initial Poisson distribution, all the Fock quanta contribute independently (i.e. they are not correlated). 

Under certain conditions, the density matrix of a reasonably general initial state can be 
represented on the coherent states basis using the Klauder-Sudarshan $P$-representation \cite{QO} and this is the reason why 
it is also interesting to examine the case of coherent states. The squeezed coherent states of relic gravitons may arise in two complementary ways mirroring their quantum optical analogs originally discussed by Caves and Yuen \cite{CY}. In the Caves representation the initial state is rotated,  squeezed and finally dispalced\footnote{We recall that  that, by definition, $|z\, \delta \rangle =  S(z) \,R(\delta) |\mathrm{vac}\rangle$ and $|\beta\rangle = D(\beta) |\mathrm{vac}\rangle$. } i.e. $|\alpha\, z\, \delta \rangle = D(\alpha) | z\, \delta \rangle$. In the Yuen representation \cite{CY} the squeezed-coherent states of relic gravitons are instead defined as $| z\, \delta\, \beta \rangle = S(z) \,R(\delta) |\beta\rangle$. Note that $D(\alpha) = e^{\alpha \hat{a}^{\dagger} - \alpha^{*} \hat{a}}$ denotes the standard Glauber displacement operator while $D(\beta)$ has the same form but a different complex amplitude. While the two approaches do not commute in general, they are fully equivalent provided $\alpha = [e^{- i \delta} \,\cosh{r}\, \beta - e^{i (\theta + \delta)} \, \sinh{r} \,\beta^{*}]$. With these notations the squeezed-coherent state is given by $| \alpha\, z\, \delta \rangle = D(\alpha) S(z) R(\delta) 
|\mathrm{vac} \rangle$, and, in this case, the associated degree of second-order coherence is:
\begin{equation}
g_{s}^{(2)} -1 = \frac{\sinh^2{r}[ \sinh^2{r} + \cosh^2{r} ] + 2 |\alpha|^2 [ \sinh^2{r} - \sinh{r} \cosh{r} \cos{ 2 \zeta}]}{[|\alpha|^2 + \sinh^2{r}]^2},
\label{IN5}
\end{equation}
where $\alpha = |\alpha| e^{i \varphi}$ and $\zeta = \varphi - \theta/2$. 
Equation (\ref{IN5}) can also be rewritten by introducing the average multiplicity of the 
coherent component $\overline{n}_{coh} = |\alpha|^2$ and of the squeezed component:
\begin{equation}
g_{s}^{(2)} -1 = \frac{\overline{n}_{sq}[ 2 \overline{n}_{sq} +1  ] + 2 \overline{n}_{coh} [ \overline{n}_{sq} - \sqrt{ \overline{n}_{sq} (\overline{n}_{sq} +1)} \cos{ 2 \zeta}]}{[ \overline{n}_{coh}+ \overline{n}_{sq}]^2}.
\label{IN6}
\end{equation}
According to Eqs. (\ref{IN5}) and (\ref{IN6}) the only region of the parameter space where $g_{s}^{(2)} < 1$ corresponds to the following pair of conditions $\overline{n}_{coh} \gg \overline{n}_{sq}$  and  $\cos{2 \zeta} >  \tanh{r}$.  Neglecting for a moment the requirements on the phases, the coherent component the demand that the coherent component greatly exceeds the squeezed contribution clashes with the assumption that the backreaction of the initial state is negligible. If the phases conspire in such a way that $\zeta \to 0$ the situation does not drastically 
improve. From Eq. (\ref{IN4}), in the case $\zeta \to 0$ the condition $[g_{s}^{(2)} -1] <0$ translates into 
$\sinh^2{r} \cosh{2 r} < |\alpha|^2 (1 - e^{-2 r}) $ that always implies $\overline{n}_{coh} \gg \overline{n}_{sq}$. 

In the Yuen representation the argument is similar to the one of Eqs. (\ref{IN5}) and (\ref{IN6}) but it is just 
obscured by a further the algebraic complication. The degree of second-order coherence to be scrutinized is now:
\begin{eqnarray}
g_{s}^{(2)} -1 &=& \frac{\overline{n}_{sq} [ 2 \overline{n}_{sq}+1 ] }{\{ \overline{n}_{sq} + \overline{n}_{coh} [ 2 \overline{n}_{sq}+1- 2 \overline{p}_{sq} \cos{2 \gamma}]\}^2}
\nonumber\\
&+& \frac{2 \overline{n}_{coh} \{ [2 \overline{n}_{sq}+1  - 2\overline{p}_{sq}  \cos{2 \gamma}] \overline{n}_{sq} - \overline{p}_{sq} [(2 \overline{n}_{sq}+1) \cos{2 \gamma} - 2 \overline{p}_{sq} ]\}}{\{ \overline{n}_{sq} + \overline{n}_{coh} [ 2 \overline{n}_{sq}+1- 2 \overline{p}_{sq} \cos{2 \gamma}]\}^2},
\label{IN5a}
\end{eqnarray}
where  the shorthand notations $\overline{n}_{sq} = \sinh^2{r}$,  $ \overline{p}_{sq} =\sqrt{\overline{n}_{sq} (\overline{n}_{sq}+1)} $ and $\overline{n}_{coh} = |\beta|^2$ have been introduced. Note that $\gamma = (\theta/2 + \delta - \chi)$ is now the overall phase and $\chi$ is the phase of the transformed coherent component (i.e. $\beta = |\beta| e^{i \chi}$). If $\overline{n}_{sq} \simeq \overline{p}_{sq} \gg 1$, Eq. (\ref{IN5a}) gives: 
\begin{equation}
g_{s}^{(2)} -1 = \frac{2 [ 1 + 4 \overline{n}_{coh}\, ( 1 - \cos{2\gamma})]}{[1 + 2 \overline{n}_{coh} ( 1 - \cos{2\gamma})]^2}.
\label{IN5b}
\end{equation}
While from Eq. (\ref{IN5b}) we conclude 
that $g_{s}^{(2)} \geq 1$ we have, in particular, that for  $\gamma = 2 \pi m$ (where $m$ is a natural number) $g^{(2)}_{s} \to 3$. Thanks to  elementary identities, Eq. (\ref{IN5a}) can also be expressed as: 
\begin{equation}
g_{s}^{(2)} -1 = \frac{\sinh^2{r} [ \sinh^2{r} + \cosh^2{r}] + \overline{n}_{coh} [ e^{- 4 r} ( 1 - e^{ 2 r})\cos^2{\gamma} + e^{4 r} ( 1 - e^{ - 2 r})\sin^2{\gamma} ] }{\{ \sinh^2{r} + |\beta|^2 [  e^{- 2 r} \cos^2{\gamma} + e^{2 r} \sin^2{\gamma} ]\}^2 }.
\label{IN5c}
\end{equation}
According to Eq. (\ref{IN5c}) the final distribution is sub-Poissonian when $e^{6 r} < 8 \overline{n}_{coh} \cos^2{\gamma}/[ 1 + 8 \overline{n}_{coh} \sin^2{\gamma}]$; if $\gamma \to 0$ the same requirement implies $r < (1/6) \ln{[ 8 \overline{n}_{coh}]}$. These conditions are incompatible with the demand that $\overline{n}_{coh} \ll \overline{n}_{sq}$ dictated by backreaction considerations\footnote{If $\gamma \neq 0$ and $1 < \overline{n}_{coh} < \overline{n}_{sq}$ we have that $r < - (1/3) \ln{[\tan{\gamma}]}$. Note that an exception to the previous statement could be represented by the case $\gamma\to 0$
which is however excluded since $\gamma$ is not bound to vanish when the modes of the field are inside the Hubble radius (see, for instance, the first and the last papers of Ref. \cite{HBT1}). }. 

All the results obtained so far in the single-mode approximation can be corroborated by the analysis 
of the full correlators in the zero-delay limit. To scrutinize  this relevant point,
the HBT correlations will be analyzed by consistently taking into account all the modes of the field for a single tensor polarization. For the sake of illustration (and to avoid a longer discussion) we shall focus the attention on an initial multiparticle Fock state $|\{ n\}\rangle$ which is also, in some sense, the most interesting choice since it minimizes the degree of second-order coherence of the relic gravitons in the single-mode case. The purpose here is to confirm that the same conclusion persists when the full spectrum of modes is taken into account and, for this end, the explicit form of the degree of second-order coherence can be written as:  
\begin{equation}
\overline{g}^{(2)}(x_{1}, x_{2}) = \frac{\langle \{ n\, \delta\, z\}|  \hat{\mu}^{(-)}(x_{1}) \, \hat{\mu}^{(-)}(x_{2})\hat{\mu}^{(+)}(x_{2}) \hat{\mu}^{(+)}(x_{1})| \{ z\,\delta\,n\} \rangle}{ \langle \{ n\,\delta\, z\}|\hat{\mu}^{(-)}(x_{1}) \, \hat{\mu}^{(+)}(x_{1})| \{ z\, \,\delta\,n\} \rangle \langle \{ n\, \,\delta \, z\}|\hat{\mu}^{(-)}(x_{2}) \, \hat{\mu}^{(+)}(x_{2})| \{ z\, \delta\, n\} \rangle},
\label{FS1}
\end{equation} 
where the single polarization of the field (\ref{exp1}) has been separated as 
$\hat{\mu}(x) = \hat{\mu}^{(+)}(x) + \hat{\mu}^{(-)}(x)$;
by definition $\hat{\mu}^{(+)}(x)$ annihilates the vacuum (i.e.  $\hat{\mu}^{(+)}(x) |\mathrm{vac} \rangle=0$);
recalling that $\hat{\mu}^{(-)}(x) = \int d^{3} k/(2\pi)^{3/2} \hat{a}_{- \vec{k}}^{\dagger}/\sqrt{2 k} e^{- i \vec{k}\cdot\vec{x}}$ we have that the one of two equivalent integrals appearing in the denominator of Eq. (\ref{FS1}) is:
\begin{eqnarray}
 \langle \{ n\, \delta\, z\}|\hat{\mu}^{(-)}(x_{1}) \, \hat{\mu}^{(+)}(x_{1})| \{ z\, \delta\, n \} \rangle= \frac{1}{2\pi^2} \int k \, d k \biggl\{ n_{k}  \, [ \overline{n}_{k}(\tau_{1})  +1] + \overline{n}_{k}(\tau_{1})\, [n_{k} +1] \biggr\}.
\label{FS2}
\end{eqnarray}
Note that $ \overline{n}_{k}(\tau_{1}) = |v_{k}(\tau_{1})|^2$ is the average multiplicity of the state 
$| \{ z\, \delta\, n \}\rangle $ whose explicit expression is:
\begin{equation} 
| \{ z\, \delta\, n \}\rangle = {\mathcal S}(z)\, {\mathcal R}(\delta)  | \{n \}\rangle, \qquad | \{n \}\rangle = \prod_{\vec{p}} \frac{\hat{a}_{\vec{p}}^{\dagger\,\,n_{p}}}{\sqrt{n_{p} !}} | \mathrm{vac}\rangle.
\label{FS2a}
\end{equation}
In Eq. (\ref{FS2a})  ${\mathcal S}(z)$ and ${\mathcal R}(\delta)$ denote the (multiparticle) squeezing and rotation operators 
\begin{eqnarray} 
&& {\mathcal S}(z) = e^{\frac{1}{2}\int d^{3} k \,[ z_{k}^{*} \, \hat{a}_{\vec{k}} 
\hat{a}_{-\vec{k}} - z_{k} \, \hat{a}^{\dagger}_{-\vec{k}} \hat{a}^{\dagger}_{\vec{k}} ]}, \qquad 
 {\mathcal R}(\delta)= e^{ - \frac{i}{2} \int d^{3}k\,\delta_{k}\, [ \hat{a}_{\vec{k}}^{\dagger} 
 \hat{a}_{\vec{k}} + \hat{a}_{-\vec{k}}  \hat{a}_{-\vec{k}}^{\dagger}]},
 \label{FS3}\\
&& {\mathcal R}^{\dagger}(\delta)\, {\mathcal S}^{\dagger}(z)\, \hat{a}_{\vec{p}}  \, {\mathcal S}(z) \, {\mathcal R}(\delta) 
= e^{- i \delta_{p}} \,\cosh{r_{p}}\,\hat{a}_{\vec{p}} - e^{ i (\delta_{p} + \theta_{p})} \,\sinh{r_{p}}\, \hat{a}_{- \vec{p}}^{\dagger},
\label{FS4}
\end{eqnarray}
where $z_{p} = r_{p}\,e^{i \theta_{p}}$. It is now convenient to use directly the parametrization 
of Eqs. (\ref{AA10}) and (\ref{AA11}) where now, according to Eq. (\ref{FS4}), $u_{p}(\tau)=  e^{- i \delta_{p}} \,\cosh{r_{p}}$ and  $v_{p}(\tau) = e^{ i (\delta_{p} + \theta_{p})}\,\sinh{r_{p}}$. Note that ${\mathcal S}(z)$ and ${\mathcal R}(\delta)$ are the multimode 
analogs of the single-mode operators $S(z)$ and $R(\delta)$ already introduced after Eq. (\ref{one}). 
After repeated use of Eq. (\ref{FS4}) all the expectation values can be explicitly evaluated so that the numerator of Eq. (\ref{FS1}) turns out to be:
\begin{eqnarray}
&& \langle \{ n\, \delta\, z\}|\hat{\mu}^{(-)}(x_{1}) \, \hat{\mu}^{(-)}(x_{2}) \, \hat{\mu}^{(+)}(x_{2}) \, \hat{\mu}^{(+)}(x_{1}) \,| \{ z\, \delta\, n \} \rangle =
\nonumber\\
&&  \int \frac{k \, d k}{4 \pi^2} \int \frac{p\, d p}{4 \pi^2} \biggl\{ \biggl[ 
n_{k} \, n_{p} \biggl( u_{k}^{*}(\tau_{1}) u_{p}^{*}(\tau_{2}) u_{k}(\tau_{2}) u_{p}(\tau_{1}) 
 + u_{k}^{*}(\tau_{1}) v_{k}^{*}(\tau_{2}) v_{p}(\tau_{2}) u_{p}(\tau_{1}) \biggr)
 \nonumber\\
&&+ \,(n_{p}+1) (n_{k}+1) \biggl(  v_{k}^{*}(\tau_{1}) v_{p}^{*}(\tau_{2}) v_{k}(\tau_{2}) v_{p}(\tau_{1})
+ v_{k}^{*}(\tau_{1}) u_{k}^{*}(\tau_{2}) u_{p}(\tau_{2}) v_{p}(\tau_{1})\biggr)
\nonumber\\
&&+\, n_{k}  (n_{p}+1) \biggl(  u_{k}^{*}(\tau_{1}) v_{k}^{*}(\tau_{2}) u_{p}(\tau_{2}) v_{p}(\tau_{1})
+ u_{k}^{*}(\tau_{1}) v_{p}^{*}(\tau_{2}) u_{k}(\tau_{2}) v_{p}(\tau_{1})\biggr)
\nonumber\\
&&+\, n_{p}  (n_{k}+1) \biggl(  v_{k}^{*}(\tau_{1}) u_{k}^{*}(\tau_{2}) v_{p}(\tau_{2}) u_{p}(\tau_{1})
+ v_{k}^{*}(\tau_{1}) u_{p}^{*}(\tau_{2}) v_{k}(\tau_{2}) u_{p}(\tau_{1})\biggr)\biggr]  j_{0}(k r) j_{0}(p r)
\nonumber\\
&&+\, (n_{k}+ 1) (n_{p}+ 1) |v_{k}(\tau_{1})|^2 \,  |v_{p}(\tau_{2})|^2 +  n_{k} n_{p} |u_{k}(\tau_{1})|^2 
\,|u_{p}(\tau_{2})|^2 
\nonumber\\
&&+\, n_{k} (n_{p}+ 1) |u_{k}(\tau_{1})|^2 \,  |v_{p}(\tau_{2})|^2  +  n_{p} (n_{k}+ 1) |v_{k}(\tau_{1})|^2 \, 
|u_{p}(\tau_{1})|^2 \biggr\}, 
\label{FS5}
\end{eqnarray}
where $j_{0}( x)= \sin{x}/x$ denotes the spherical Bessel function of zeroth order resulting from the 
angular integration.
Note that if the average multiplicity of the initial state is set to zero for all the modes (i.e. $n_{k} = n_{p} =0$) the HBT correlations reproduce the case of relic gravitons produced  from the vacuum.
From Eqs. (\ref{FS2}) and (\ref{FS5}) we can explicitly compute the degree of second-order coherence in the zero-delay limit and in the physical case of large average multiplicities of the produced particles:
\begin{equation}
\lim_{x_{1} \to x_{2} } g^{(2)}(x_{1}, x_{2}) =   \frac{12}{8} \frac{\int k \,d k \,\int p \, d p\, n_{k} \, n_{p}\, \overline{n}_{k}(\tau_{1}) \, 
\overline{n}_{p}(\tau_{1}) }{\int k \,d k \, n_{k} \, \overline{n}_{k}(\tau_{1}) \, \int p \,d p \,n_{p}\, \overline{n}_{p}(\tau_{1})} \to \frac{3}{2},
\label{FS6}
\end{equation}
where the multiplicity of the initial state has been kept fixed. The result of Eq. (\ref{FS6}) coincides with the 
degree of coherence obtained in the single-mode approximation. The second equality in Eq. (\ref{FS6}) follows by computing the iterated integral in the numerator and by noting that
it coincides with the result of the denominator. 

All in all the degree of second-order coherence of the relic gravitons is always super-Poissonian and this general conclusion is not altered if the production of the quanta takes place from the vacuum or from a generic initial state that does not minimize the tensor Hamiltonian. The quantitative expressions of the final HBT correlations are  however sensitive to the statistical properties of  the different initial states provided the total number of inflationary efolds is not too large. Denoting with $\overline{g}^{(2)}$ the degree of second-order coherence of the relic gravitons in the case of a single tensor polarization, the present analysis supports the validity 
of the following general bound:
\begin{equation}
\frac{3}{2} \leq \overline{g}^{(2)} \leq 3,
\label{FS7}
\end{equation}
where $3/2$ corresponds to the minimal value associated with an initial Fock state while $3$ is the value of the squeezed vacuum state. The range of Eq. (\ref{FS7}) always exceeds $1$, which is the result valid for a coherent state or for a mixed state with Poissonian weights. Furthermore if a mixed state is characterized by the Bose-Einstein statistical weights, the degree of second-order coherence goes to $2$. 
The multiplicity distribution of the gravitons produced from the vacuum is also of Bose-Einstein type but with a larger amount of correlations (hence the value $3$ instead of $2$ for the normalized degree of second-order coherence) \cite{HBT1}.  However the statistics of the relic gravitons is super-chaotic (i.e. larger than in the Bose-Einstein case) either when the initial state coincides with the vacuum or when the total duration of inflation 
greatly exceeds the critical number of efolds [i.e. about ${\mathcal O}(66)$ for the fiducial set of parameters of the concordance paradigm]. The presence of initial states different from the vacuum may reduce the degree of second-order coherence; this effect is conceptually interesting but practically only relevant when the total number of inflationary efolds gets very close to its critical value. We examined a variety of initial states first in the single-mode approximation and then for a continuous set of modes. It turns out that if the initial state has a sub-Poissonian statistics (as in the case of multiparticle Fock states) the degree of second-order coherence of relic gravitons produced by stimulated emission is always super-Poissonian but it can be smaller than $3$. The conclusions of the present analysis hold as long as the average multiplicity of the gravitons produced from the vacuum  is much larger than the average multiplicity of the initial state. If the degree of quantum coherence of the relic gravitons will be one day  accessible to direct interferometric measurements of HBT type, then Eq. (\ref{FS7}) could be used to disambiguate the physical origin of the signal.

The author wishes to thank T. Basaglia, A. Gentil-Beccot, M. Medves and S. Rohr of the CERN Scientific Information 
Service for their assistance.


\newpage

\begin{thebibliography}{99}

\bibitem{sakharov} A. D. Sakharov, Zh. Exp. Teor. Fiz. {\bf 49}, 345 (1965) [Sov. Phys. JETP {\bf 22}, 241 (1966)].

\bibitem{GR}  L.~P.~Grishchuk,   Sov.\ Phys.\ JETP {\bf 40}, 409 (1975)   [Zh.\ Eksp.\ Teor.\ Fiz.\  {\bf 67}, 825 (1974)];  Annals N.\ Y.\ Acad.\ Sci.\  {\bf 302}, 439 (1977);   L. H. Ford and L. Parker, Phys. Rev. {\bf D16}, 1601 (1977); Phys.\ Rev.\ D {\bf 16}, 245 (1977).

\bibitem{INF} A.~A.~Starobinsky, JETP Lett.\  {\bf 30}, 682 (1979) [Pisma Zh.\ Eksp.\ Teor.\ Fiz.\  {\bf 30}, 719 (1979)];
V. A. Rubakov, M. V. Sazhin and A. V. Veryaskin, Phys. Lett. {\bf 115B}, 189 (1982).

\bibitem{RT} P.~A.~R.~Ade {\it et al.} [BICEP2 and Keck Array Collaborations], Phys.\ Rev.\ Lett.\  {\bf 116}, 031302 (2016).

\bibitem{INTER} J.~Aasi {\it et al.},  Phys.\ Rev.\ Lett.\  {\bf 113},  231101 (2014).
 B.~P.~Abbott {\it et al.},  Phys.\ Rev.\ Lett.\  {\bf 118}, 121101 (2017)
  Erratum: [Phys.\ Rev.\ Lett.\  {\bf 119},  029901 (2017)].

\bibitem{QO} L. Mandel and E. Wolf, {\it ``Optical coherence and quantum optics''}, 
(Cambridge University Press, Cambridge, 1995).

\bibitem{HBT0} R. Hanbury Brown and R. Q. Twiss, Nature {\bf 178}, 1046 (1956); 
 Proc. Roy. Soc. (London) {\bf A242}, 300 (1957); 
Proc. Roy. Soc. (London)  {\bf A243}, 291 (1958).

\bibitem{HBT1} M.~Giovannini,  Phys.\ Rev.\ D {\bf 83}, 023515 (2011);  
Class.\ Quant.\ Grav.\  {\bf 34},  035019 (2017); Mod.\ Phys.\ Lett.\ A {\bf 32},  1750191 (2017); S.~Kanno and J.~Soda,  arXiv:1810.07604 [hep-th]; M.~Giovannini,  arXiv:1902.11075 [hep-th].

\bibitem{glauber} R.~J.~Glauber,  Phys.\ Rev.\ Lett.\  {\bf 10}, 84 (1963); E. C. Sudarshan,  Phys.\ Rev.\ Lett.\  {\bf 10}, 277 (1963).

\bibitem{NN}  A. R. Liddle and S. M. Leach, Phys. Rev. D {\bf 68},  103503 (2003); M.~Giovannini,
  Class.\ Quant.\ Grav.\  {\bf 20}, 5455 (2003); Phys.\ Rev.\ D {\bf 88},  021301 (2013);  Phys.\ Rev.\ D {\bf 89}, 123517 (2014).

\bibitem{CY} C.~M.~Caves, Phys. Rev. D  {\bf 23}, 1693 (1981); H.~P.~Yuen, Phys.\ Rev.\  A {\bf13}, 2226 (1976).
\end{thebibliography}
\end{document}